\documentclass{ametsocV6.1}
\nolinenumbers

\makeatletter
\@ifundefined{internallinenumbers}{%
    \newcommand{\internallinenumbers}{\relax}% Define it to do nothing
}{%
    \renewcommand{\internallinenumbers}{\relax}%
}
\makeatother

\title{CloudCast – Total Cloud Cover Nowcasting with Machine Learning}

\authors{Mikko Partio\aff{1}\aff{2},\correspondingauthor{Mikko Partio, mikko.partio@fmi.fi} 
Leila Hieta\aff{1}, 
Anniina Kokkonen\aff{1}
}

\affiliation{
{\aff{1}Finnish Meteorological Institute}\\
{\aff{2}Aalto University}\\
}

\abstract{
Cloud cover plays a critical role in weather prediction and impacts several sectors, including agriculture, solar power generation, and aviation. Despite advancements in numerical weather prediction (NWP) models, forecasting total cloud cover remains challenging due to the small-scale nature of cloud formation processes. In this study, we introduce CloudCast, a convolutional neural network (CNN) based on the U-Net architecture, designed to predict total cloud cover (TCC) up to five hours ahead. Trained on five years of satellite data, CloudCast significantly outperforms traditional NWP models and optical flow methods. Compared to a reference NWP model, CloudCast achieves a 24\% lower mean absolute error and reduces multi-category prediction errors by 46\%. The model demonstrates strong performance, particularly in capturing the large-scale structure of cloud cover in the first few forecast hours, though later predictions are subject to blurring and underestimation of cloud formation. Model selection was performed using sensitivity experiments which identified the optimal input features and loss functions, with MAE-based models performing the best. CloudCast has been integrated into the Finnish Meteorological Institute’s operational nowcasting system, where it improves cloud cover forecasts used by public and private sector clients. While CloudCast is limited by a relatively short skillful lead time of about three hours, future work aims to extend this through more complex network architectures and higher-resolution data. CloudCast code is available at \url{https://github.com/fmidev/cloudcast}.
} 

\begin{document}

\maketitle

\section{Introduction}

Clouds play a vital role in Earth’s atmospheric energy balance by regulating temperature, distributing moisture, and influencing weather patterns \citep{Quante2004}. Their effects extend across various sectors, including agriculture, solar power generation, aviation, and everyday life, where cloud cover forecasts help people plan activities \citep{Lazo2009}. However, predicting cloud cover accurately remains a challenge for numerical weather prediction (NWP) models, due to the small-scale nature of cloud formation \citep{Tiedtke1993}, the spatial and temporal variability of observational data \citep{Dai2006}, and the complexities of assimilating remote-sensing data into NWP models \citep{Bauer2011}. Nowcasting, defined as short-term forecasting of up to 6 hours ahead \citep{Wang2017}, relies on frequently updated, high resolution observational data, which is typically extrapolated into short range forecasts using methods such as optical flow and neural networks. While precipitation nowcasting using radar information is well-established and traditional methods provide accurate forecasts up to three hours, neural networks have extended this horizon to 12 hours \citep{Sonderby2020}. However, cloud cover nowcasting is less developed because of the lack of high-quality observational data that directly describe cloud fraction. Cloud fraction cannot be easily derived from raw satellite measurements.

Optical flow methods use sequences of images, such as precipitation maps, to estimate the movement of the field. They typically operate in two phases: first, consecutive frames of cloud cover are analyzed to generate an atmospheric motion vector (AMV) field, which describes the past movement of weather phenomena. This AMV field is applied to the current frame to extrapolate future states. The process is repeated, using each new forecast as input, but without recalculating the AMV field until the desired forecast length is achieved. Since these methods only advect the existing field and do not account for the growth or decay of clouds, their prediction horizon is limited to a few hours. Operational forecasts often combine optical flow methods with numerical weather prediction (NWP) models to produce seamless forecasts. Typically, the first one or two hours rely on optical flow methods. As the forecast quality from optical flow begins to decline, NWP data is used to continue the forecast \citep{Blahak2021, Roberts2023}. Machine learning methods can learn complex, nonlinear cloud dynamics, enabling them to capture both advection and formation/dissipation processes, whereas optical flow primarily models motion based on pixel displacement. Additionally, ML models can leverage historical patterns and multi-source data (e.g., satellite, reanalysis) to improve forecasts, while optical flow relies solely on recent frame sequences, limiting its ability to predict cloud evolution beyond short lead times.

In machine learning, convolutional neural networks (CNNs) use a hierarchical structure where information content is increasingly abstracted as the data is passed through the network. Typical models are autoencoders that contain a contracting path that captures high-level features and an expanding path that reconstructs the output. CNNs are also becoming increasingly popular for sequence-to-sequence nowcasting tasks, such as predicting precipitation and cloud cover \citep{Moskolai2021}. Various architectures have been successful in different studies, most commonly using either explicit-memory models (LSTM, GRU) or architectures where temporal relations are described implicitly, such as U-Net. \citet{Feng2022} developed a shallow CNN to predict solar irradiance from sky imager data. \citet{Berthomier2020} trained several different CNNs, including a U-Net to predict a binary cloud mask, outperforming NWP forecasts. Similarly \citet{Lagerquist2021} used a U-Net to predict convection using multipectral images from Himawari-8, and \citet{Fan2024} used a ResNet (residual neural network) to predict convection initialization from GOES-16 satellite observations. CNNs have also been combined with physics-informed modules \citep{Montassir2023} as well as optical flow methods \citep{Ritvanen2023}. In time series prediction, recurrent neural networks (RNN) have been utilized widely. These networks are designed to learn temporal dependencies in the data, with network layers that can model the temporal dimension present in the data. For spatio-temporal prediction tasks typically either LSTM \citep{Nielsen2021, Shi2015, Kumar2020, Xia2024} or GRU \citep{Kellerhals2022, Wang2023, Leinonen2022} based networks are used. The inclusion of memory layers in RNN-based networks increases their complexity and makes them more difficult to train compared to similar architectures that do not incorporate memory.

This study introduces CloudCast, a deep convolutional neural network for total cloud cover (TCC) prediction based on the U-Net architecture and trained on five years of satellite-based cloud cover data. CloudCast produces a five-hour forecast that outperforms both optical flow methods and NWP in our benchmarks. 

Finnish Meteorological Institute (FMI) provides an hourly-updated, 10-day forecast used by other government agencies, private sector and public. The forecast is a combination of direct model output from numerical weather prediction (NWP) models, statistically calibrated data and observation extrapolation data. CloudCast is integrated to FMI's operational weather production and hourly forecasts are available to all FMI users.

The rest of the paper is organized as follows: Section 2 describes the curation of training and validation data, the selection of the neural network model, the sensitivity experiments, and the model training details. Section 3 presents the results from the sensitivity experiments and the verification results for CloudCast, including case studies. Finally, section 4 concludes the study and outlines directions for future research.

\section{Methods}

This section formulates our ML approach by first discussing the dataset used for model training and selection (see Section 2a). Then the proposed neural network model is described in Section 2b. Finally, we discuss the loss functions and hyper-parameters used for model training in Section 2c.

\subsection{Data}

\begin{figure}[h]
 \centerline{\includegraphics[width=34pc]{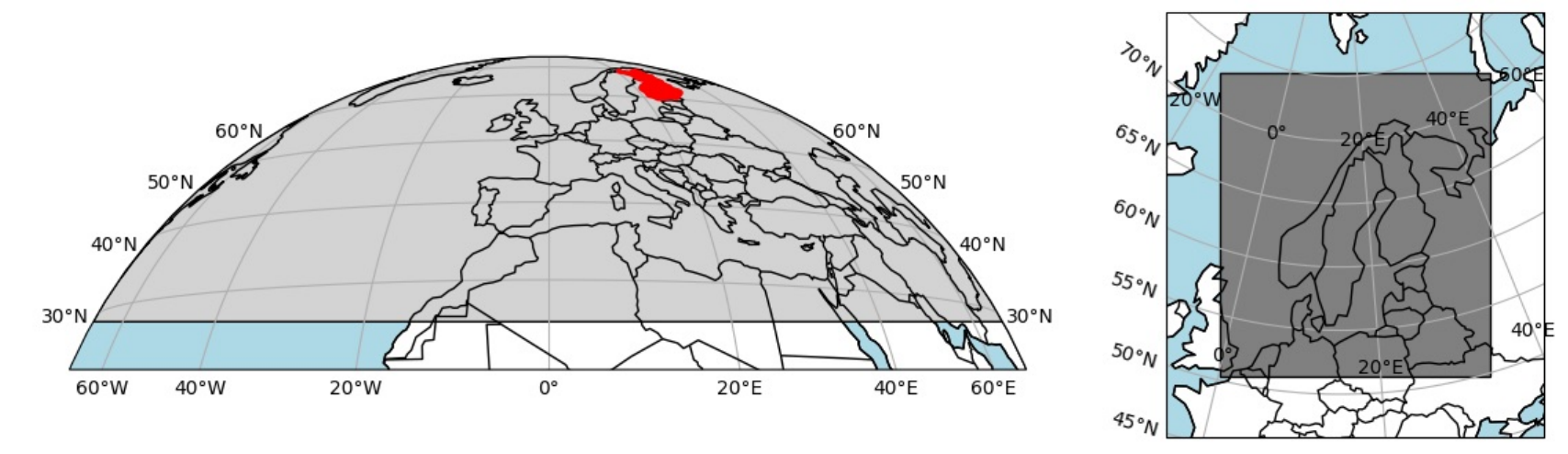}}
  \caption{Left: Northern half-hemisphere as produced by FMI's NWCSAF installation. Finland highlighted in red. Right: CloudCast domain covering Northern Europe in Lambert conformal conic projection}\label{geographical_domains}
\end{figure}

Meteosat-10 is a geostationary satellite operated by EUMETSAT and located above \text{0\textdegree} latitude, \text{0\textdegree} longitude. It provides a full-disk image every 15 minutes and has a spatial resolution of 3km at nadir. Due to its position over the equator, the field of vision narrows significantly towards the north. The effective resolution in latitudes over \text{60\textdegree} can be up to 12km. Finland is located between latitudes \text{59\textdegree} and \text{70\textdegree} North (Figure \ref{geographical_domains}).

Spinning Enhanced Visible and Infrared Imager (SEVIRI), an instrument onboard Meteosat-10, provides data in 12 spectral bands \citep{Aminou1997}. This data, along with any required auxiliary data, is processed with NWCSAF software suite for geostationary satellites (NWC/GEO) \citep{Kerdraon2021}. As a result a direct estimate of the total cloud cover, a parameter called effective cloudiness, is produced. Effective cloudiness is a part of the cloud top temperature and height (CTTH) family and is defined as the cloud fraction multiplied by cloud emissivity in the 10.8µm window channel \citep{Kerdraon2021}. It has a value of 1 for thick clouds, 0 for cloud-free areas and a value between 1 and 0 for semi-transparent clouds. The parameter has been empirically determined to closely correspond to the TCC observed and forecasted (by NWP). Effective cloudiness is produced in real-time in 15-minute intervals as a part of the FMI operational systems supporting duty forecasters. 

\begin{figure}
\center
\includegraphics[width=29pc]{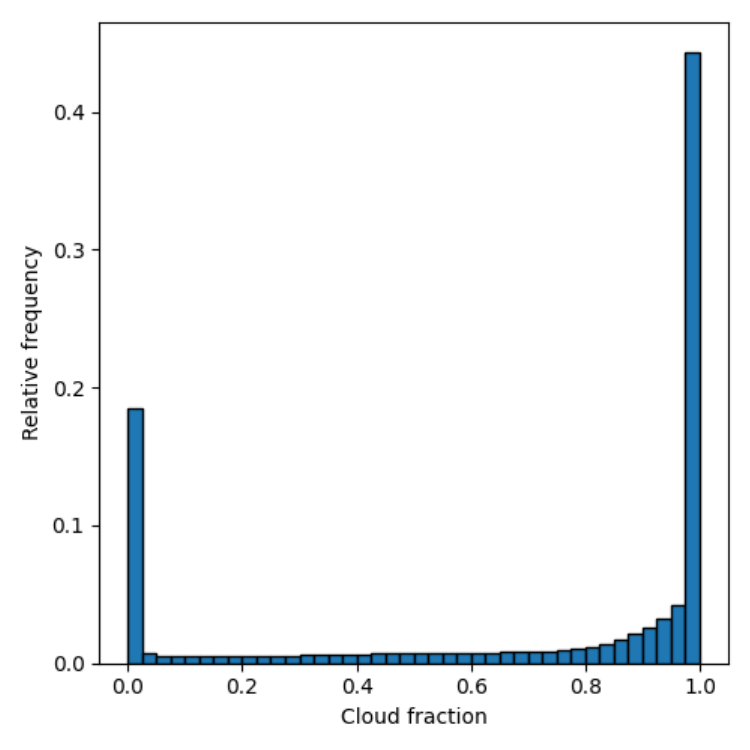}
\caption{Histogram showing the distribution of effective cloudiness in the training dataset for Scandinavia from November 2018 to October 2023.}\label{histogram}
\end{figure}

Effective cloudiness data from November 2018 to October 2023 was used for training, a total of 1795 days and 166574 individual images. For evaluating the model, data from November 2023, January, April and June 2024 was used, totalling to 121 days and 11598 images and covering all seasons. FMI archive for effective cloudiness data extends over three different NWC/GEO versions: v2016, v2018 and v2021; no significant change in the meteorological quality of the data between versions was found. There were gaps in the time series, due to issues either at EUMETSAT or FMI production systems, but they were not addressed in any way. The FMI data archive is quality-controlled, and aside from gaps in the time series, no traces of data corruption or other anomalous values were discovered. The histogram of the cloud cover training data is shown in Figure \ref{histogram}. The cloud cover data have a strong bimodal distribution, where more than 60\% of the data is either zero or one.

NWCSAF software outputs the full disk data in NetCDF format in a geostationary satellite specific spatial projection. The data was reprojected to lambert conformal conic projection matching the domain of interest, and converted to grib file format. This geographical domain is located between 50\textdegree N and 75\textdegree N, 20\textdegree W and 45\textdegree E, as seen in Figure \ref{geographical_domains}. The horizontal resolution of the data after reprojection was 4 km. In high latitudes the effective resolution is worse than this as the satellites view is less orthogonal with relation to the surface of the Earth. A parallax-corrected version of the data was used to reduce the distortion in the data. All missing values were replaced with zero and the data was scaled between zero and one.

The method developed in the study was benchmarked against the conventional methods. The first benchmark data is the output from the control member from MetCoOp Ensemble Prediction System (MEPS). MEPS is using Harmonie-AROME cycle 43h2.2 \citep{Bengtsson2017}, a non-hydrostatic, convection-permitting numerical weather prediction system run in cooperation by the meteorological institutes of Estonia, Finland, Norway and Sweden \citep{Muller2017}. MEPS forecast is produced every three hours and it provides an hourly forecast up to 66 hours, with a horizontal resolution of 2.5km. The cloud fraction is derived for each of the 65 vertical levels of the model using a statistical cloud scheme, and the total fraction is computed assuming a random overlap between successive layers. For verification the analysis field and the first five hours of the forecast were used. The second benchmark data was EXIM, an optical flow prediction system which is a part of the NWCSAF suite of products and (in FMI's configuration) provides a 60-minute forecast in 15-minute prediction intervals \citep{ZAMG2021}. The horizontal resolution of EXIM is the same as the ground truth data, 3km. FMI archives contain EXIM data only for April and June 2024. The third benchmark data was Eulerian persistence: using the current cloudiness data as a (static) prediction field for all future predictions, which especially for the first intra-hour forecasts can be hard to beat.

\subsection{Neural network}

The CloudCast architecture was inspired by the U-Net model originally developed for image segmentation \citep{Ronneberger2015}. U-Net is an encoder-decoder convolutional neural network named after its symmetric U-shaped structure, which features a contracting path to capture context and an expanding path for localization. It employs skip connections, allowing input data to be directly transferred from the encoder to the decoder. Although initially developed for medical imaging, U-Net has proven to be a simple yet effective network, successfully applied to various domains. The advantages of U-Net over alternative convolutional networks include: 1) fast training due to its end-to-end structure and context-based learning, 2) skip connections that enable the reuse of low-level features, improving accuracy without requiring large datasets, and 3) the ability to capture both global context (via the encoder) and fine pixel-level details (via the decoder) \citep{Siddique2021}. U-Net has been widely adopted for many spatio-temporal and environmental forecasting tasks, often outperforming models with explicit temporal encoding (such as LSTMs or GRUs), as discussed in Section 1. This motivated the choice of U-Net, as it effectively captures spatial patterns while leveraging temporal context through multi-scale feature extraction. Furthermore, the relative simplicity of the U-Net makes it easier and faster to train compared to, for example, Generative Adversarial Networks (GANs) or diffusion-based networks. Given that the amount of satellite-based ground truth data available is limited to past six years, it is crucial to consider whether sufficient data exists to effectively train the network. Convolutional networks, which benefit from the strong inductive bias introduced by their convolutional layers, generally require less data than attention-based networks such as Vision Transformers \citep{Cao2022}. The U-Net architecture was implemented in CloudCast using the TensorFlow framework \citep{Abadi2016}.

\begin{figure}
\center
\includegraphics[width=39pc]{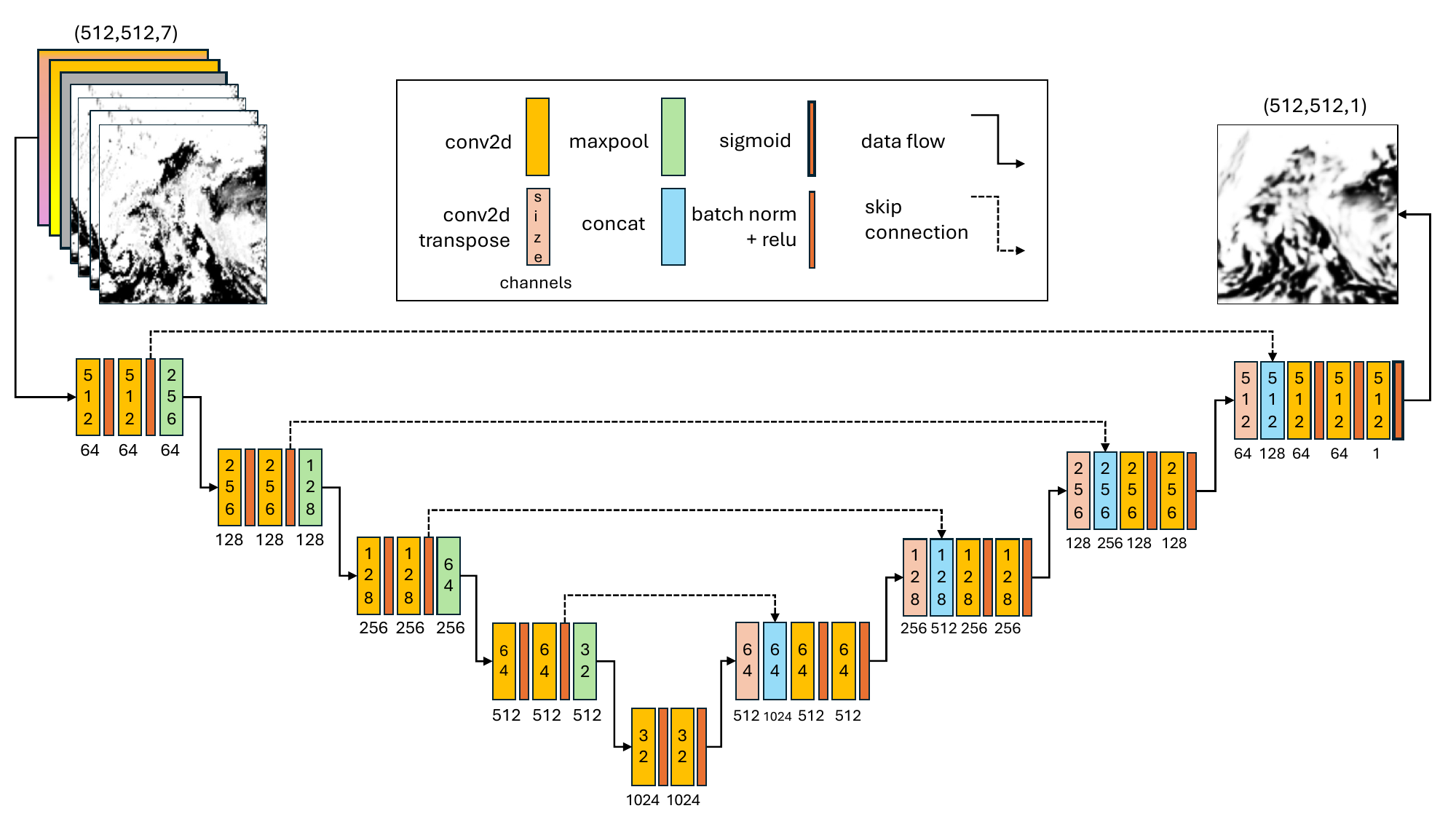}
\caption{Illustration of the CloudCast architecture. Layout follows the usual U-Net shape where input data is contracted in the encoder (left hand side) and subsequently expanded in the decoder (right hand side). The model’s input consists of seven channels: four consecutive effective cloudiness images, lead time, and date-time information. The output is a single-channel image predicting the cloudiness value.}\label{unet}
\end{figure}

The network architecture is shown in Figure \ref{unet}. The input comprises a sequence of effective cloudiness images, along with lead time and date/time information. As the data passes through the convolutional and max-pooling layers in the encoder, its resolution is gradually reduced, allowing for the extraction of increasingly abstract, high-level features. The encoder consists of four consecutive stages, during which the spatial dimensions are reduced to 32x32 pixels while the number of channels increases to 1024. Each stage includes convolutional layers followed by an activation function and a max-pooling layer to reduce spatial dimensions. The backbone of the network—situated between the encoder and decoder—also comprises convolutional layers followed by an activation function. The decoder mirrors the encoder’s structure but replaces convolutional layers with Conv2DTranspose layers to upsample the dense latent representation gradually back to the original spatial resolution. Additionally, the input data is concatenated directly with the decoder path to ensure the preservation of high-level features in the output. The final layer applies a sigmoid activation function to constrain the output values between zero and one, yielding a pixel-wise prediction of cloud cover for the specified lead time. The model comprises 31 million trainable parameters.

The approach of \citet{Sonderby2020} was adopted, where the forecast lead time is concatenated with the input data sequence. This method allows each lead time to be forecasted independently, enabling parallelization of the entire process and ensuring that each prediction is conditioned solely on ground truth data. However, using lead time as an additional input feature has drawbacks: the maximum lead time becomes an input parameter that must be determined prior to training the model, and the model’s ability to predict further into the future is constrained, as it does not observe the evolution of the weather system. While an autoregressive model can theoretically generate infinitely long predictions, the temporal range for this model is intentionally limited. This is acceptable as the focus of this study is exclusively on the nowcasting range of up to five hours.

\subsection{Training details}

To determine the optimal loss function and set of input features, systematic sensitivity experiments were conducted. These experiments tested different numbers of sequential input images—1, 2, 4, and 6—representing temporal histories of 0, 15, 45, and 75 minutes, respectively. Additionally, various static datasets were incorporated, including topography, ground type, date and time, and sun elevation angle. The strong bimodal distribution of the cloud cover data made selecting the right loss function particularly crucial: a total of six different loss functions were tested. 

Some of the widely used loss functions in machine learning, such as mean absolute error (MAE) and mean squared error (MSE), may not perform well with cloud cover, as they tend to minimize validation error by consistently predicting 50\% cloud cover. They still serve as a very strong base line for the other loss functions. Binary cross-entropy (BCE) measures the difference between predicted and actual probability distributions and is typically used in binary classification tasks to predict the probability that an input belongs to one of two categories. In the case of CloudCast, the prediction is a two-dimensional array of cloud fraction values between zero and one. Notably, the distribution of these values is heavily concentrated at zero and one (see Figure \ref{histogram}). Because the cloud fraction values predominantly resemble a binary distribution, BCE can be used as a loss function, despite the non-categorical nature of the cloud cover data. \citet{Jungersen2022} found in their study BCE+MAE, a combination of BCE and MAE, to outperform plain BCE. The log-cosh function computes the logarithm of the hyperbolic cosine of the prediction error. It behaves like MSE for small error but is robust against outliers like MAE, making it suitable for time series prediction where outliers are present \citep{Saleh2022}. Lastly, the Structural Similarity Index Measure (SSIM) assesses the similarity between two images based on structure, brightness, and contrast. It computes scores for multiple patches using a sliding window mechanism and aggregates these to provide a single global value for the image pair \citep{Wang2004}. As a loss function, SSIM evaluates structural similarity in images, emphasizing larger-scale spatial patterns rather than pixel-wise errors like MAE.

The models trained in the sensitivity experiments are summarised in Table \ref{t1}. Date and time were converted into cyclic variables using sine and cosine transformation. The terrain type was consolidated from 30 different categorical values into 6 major types: forest, urban, bare, permanent snow, water, agriculture, and other. The sun elevation angle was precomputed for each grid point and each hour of the year using an internal formula from FMI. Model names are abbreviated using convention loss function - number of input frames - additional features. Due to limited computation budget, the models were trained in a horizontal resolution of 20km.

\begin{table}[h]
\caption{Models included in the sensitivity experiments. A total of 13 different models were trained. Model names are abbreviated using convention \texttt{loss function-number of input frames-additional features}, where the additional features are \texttt{nd}: no additional features, \texttt{dt}: date and time, \texttt{terrain}: terrain type, \texttt{topo}: topography and \texttt{sun}: sun elevation angle.}
\label{t1}
\begin{center}
\begin{tabular}{ccccrrcrc}
\topline
$Name$ & $Loss function$ & $Inputs$ & $Additional features$ \\
\midline
bc-4-nd & binary crossentropy & 4 & \\
bcl1-4-nd & bc + mae & 4 & \\
mae-1-nd & mae & 1 & \\
mae-2-nd & mae & 2 & \\
mae-4-nd & mae & 4 & \\
mae-4-dt & mae & 4 & date and time \\
mae-4-terrain & mae & 4 & terrain type \\
mae-4-topo & mae & 4 & topography \\
mae-4-sun & mae & 4 & sun elevation angle \\
mae-6-nd & mae & 6 & \\
mse-4-nd & mse & 4 & \\
logcosh-4-nd & log cosh & 4 & \\
ssim-4-nd & structural similarity index & 4 & \\
\botline
\end{tabular}
\end{center}
\end{table}

The training data was converted from GRIB files into NumPy arrays as 32-bit floating-point values, totaling 11 GB. The arrays were uncompressed, allowing the data be accessed efficiently using memory mapping without needing to load it all into memory. Model training was performed with 16-bit precision. Adam optimizer \citep{Kingma2015} was used with default parameters: a learning rate of 0.001, $\beta_1$ = 0.9, and $\beta_2$ = 0.999. The batch size was set to 32. Training continued until there was no improvement in the validation loss for seven epochs, with a threshold of 0.001. The learning rate was reduced by an order of magnitude after encountering a validation loss plateau for 5 epochs. The training data set was partitioned into sequences of 24 consecutive images, comprising four historical images and 20 target images for prediction. These sequences were then shuffled and split into training and validation sets with an 80/20 ratio. Training was conducted on an Nvidia V100 GPU with 32 GB of memory and CUDA version 11.8. On average, training each model took four hours.

The best model was selected based on the results of the sensitivity experiments and retrained using a 5km horizontal resolution (corresponding to 512x512 image size) over the same time period. The dataset size was 166GB. All the model hyperparameters were kept same, except batch size, which was reduced to 4. The training time for this model was 6 days.

\section{Results}

In this section we describe the results of the sensitivity experiments (Section 3a), that were used to select the final model for further retraining. These results of the final model are discussed in Section 3b, and three case studies are presented in Section 3c. Finally, the operational application is described briefly in section 3d.

\subsection{Sensitivity experiments}

A total of 13 different models were trained to identify the optimal combination of loss function and input features. Effective cloudiness itself was used as the ground truth in the evaluation. Verification was conducted using a dataset that the model had not been exposed to during training. This dataset covers November 2023, as well as January, March, and June of 2024. Cloud cover attributes and satellite sensing capabilities vary between seasons, so using different seasons for verification provides insights into the model’s performance throughout the year.

To evaluate and score the models three different verification scores were used: mean absolute error (MAE), chi-squared value and power spectral density (PSD). The final score was the mean of the individual scores over grid and all the lead times. Chi-squared test compares the observed and expected frequencies across the dataset. The data was categorized into 20 bins and the chi-squared statistics for each model were compared to determine which models’ forecasts distribution aligns most closely with the ground truth. PSD was used to quantify the energy contained at different spatial scales, helping eliminate models that fail to capture small-scale features. Scale-separation was done using discrete fourier transform where the power of each frequency (scale) was the squared magnitude of the coefficients. Hanning windowing was used to prevent spectral leakage that otherwise would distort the results for the smallest and largest scales. Power density values were averaged across scales, with each scale weighted by the inverse of its length to emphasize the smaller scales more strongly.

\begin{table}[h]
\caption{Results from the sensitivity experiments, sorted from best to worst. Each verification score was derived from the test dataset covering four months. Models are ranked from best to worst (best=1, worst=15) ie. smaller value indicates better performance. The total rank is derived by summing of the points from the three different tests and sorting from smallest to largest. The best model for each category is underlined.}\label{t2}
\begin{center}
\begin{tabular}{ccccrrcrc}
\topline
$Model$ & $MAE$ &  $PSD$ & $Chi-squared$ & $Rank$\\
\midline
mae-4-dt	& 2	& 2 & 2 & \textbf{\underline{1}} \\
mae-6-dt    & \textbf{\underline{1}} & 3 & 5 & 2 \\
mae-2-dt    & 6 & 4 & \textbf{\underline{1}} & 3 \\
mae-4-sun	& 3 & 5 & 7 & 4 \\
mae-6-nd	& 5 & 6	& 6 & 5 \\
mae-4-topo	& 3	& 7	& 8 & 6 \\
mae-2-nd	& 9	& 8 & 3 & 7 \\
mae-4-nd	& 6	& 10 & 4 & 7 \\
mae-4-terrain & 6 & 9 & 10 & 9 \\
ssim-4-nd	& 15 & \textbf{\underline{1}} & 11 & 10 \\
bcl1-4-nd	& 10 & 12  & 12 & 11 \\
mae-1-nd	& 14 & 11 & 9 & 11 \\
bc-4-nd	&  11 & 14 & 13 & 13 \\
logcosh-4-nd & 12 & 13 & 14 & 14 \\
mse-4-nd	& 13 & 15 & 15 & 15 \\
\botline
\end{tabular}
\end{center}
\end{table}

Results of the sensitivity experiments are presented in Table \ref{t2}, ranking the models from best to worst. For brevity, only the total score over all seasons is shown. Models trained with MAE as loss function performed better than those with other types of loss functions. This can be partly attributed to the implicit advantage these models had, as they were trained and evaluated using the same metric. Nevertheless, these models also performed well on the other two tests. The model using MSE as loss function had most detail in the small scale structures, as indicated by the smallest value in PSD test, but at the same time it was the worst in chi-squared test, failing to reproduce the distribution of the ground truth. Based on the combined scores from the three tests, five out of the six best-performing models used MAE as the loss function and used four input images. The overall best-performing model was \textit{mae-4-dt}, which besides the four input images also uses date and time information as predictors. This model was selected for further retraining in horizontal resolution of 5km.

\begin{figure}
\center
\includegraphics[width=39pc]{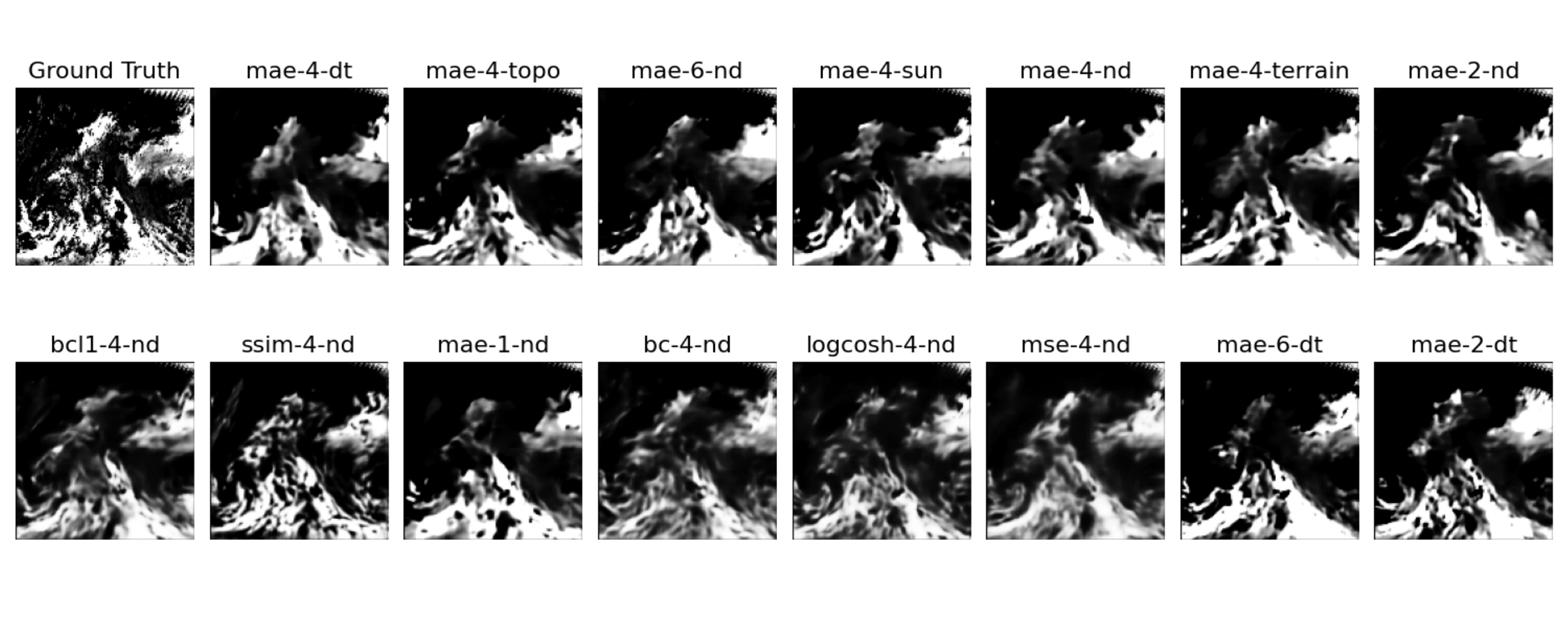}
\caption{Five-hour prediction from each of the models and the corresponding ground truth. Models are sorted from best to worst based on overall verification score. Forecast initialisation time is June 17 2024, 1200 UTC.}\label{case-128}
\end{figure}

Figure \ref{case-128} presents a five-hour forecast from each model alongside the ground truth. Clear skies are shown in white, while overcast conditions are depicted in black. As expected, all forecasts appear blurrier compared to the ground truth. The models using binary cross-entropy, log-cosh, and MSE exhibit more broken cloud cover (grey areas) than the others. The model trained with the SSIM loss function stands out with high contrast, displaying good 0/1 separation, which is ideal; however, the quantitative verification scores did not support using SSIM as a loss function. A visual artifact in the top-right corner is due to limitations in the satellite instrument’s viewing angle.

\subsection{Full resolution model}

Based on the results of the sensitivity experiments MAE was chosen as the loss function of CloudCast. The number of input images was four, and date and time were used as input features.
 
A useful forecast is one that has skill. To remove the possibility that a forecast is good by chance, the definition of a skillful forecast was adopted from \citet{Wilks2011}: the forecast outperforms a reference forecast, such as climatology or random forecast. \citet{Hogan2009} introduced mean absolute error skill score (MAESS) for continuous cloud cover data. MAESS compares a forecast against a reference forecast, typically either a random forecast or the climatological value. In this case, climatology was used: for each of the four verification months, the mean value was calculated for each grid point and used as the reference forecast. MAESS values range from zero or less (no skill, equal to or worse than climatology) to one (perfect forecast). 

\begin{figure}
\center
\includegraphics[width=39pc]{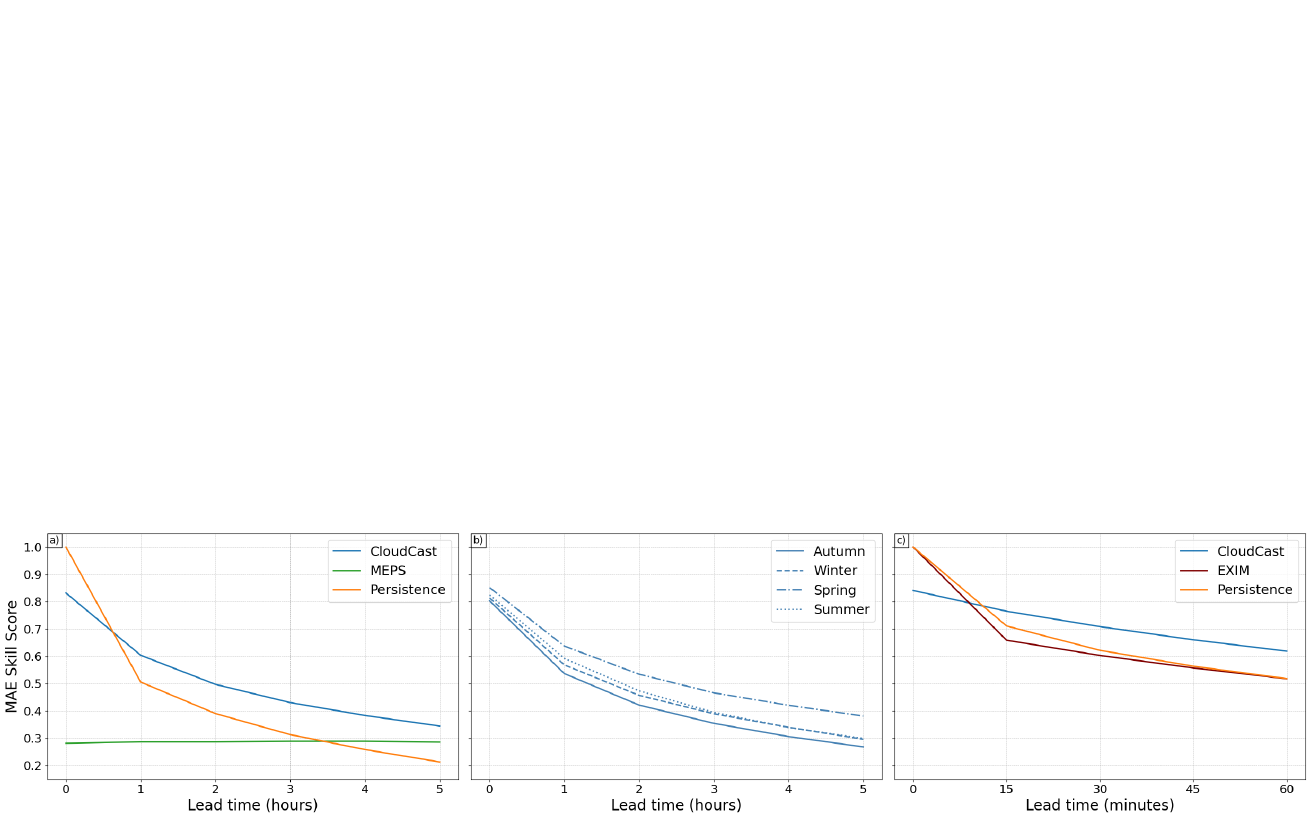}
\caption{Mean Absolute Error Skill Score as a function of lead time for a) seasonal average for a five hour forecast, b) season-wise scores for CloudCast only, and c) intra-hour prediction for the first hour. Higher values indicate better performance.}\label{f-maess}
\end{figure}

The models and ground truth each had different horizontal resolutions. We performed nearest point interpolation to match the models to the ground truth. For MEPS, we upscaled the data from 2.5km to 3km, for cloudcast we downscaled from 5km to 3km. EXIM data was already produced in the same resolution as ground truth.
MAESS was computed separately for each season and across all seasons. Figure \ref{f-maess}a shows MAESS as a function of lead time, averaged over all seasons (spring: MAM, summer: JJA, autumn: SON, winter: DJF). The persistence model achieves a perfect score for the 0-hour prediction, since it represents the last observed state before forecasting begins. However, its accuracy drops to half within the first hour and continues to decline as the forecast lead time increases. The MEPS model shows a consistently low MAESS value of around 0.3, indicating that the forecast skill is independent of the lead time.
NWP models merge observational inputs, which are often spatially sparse and temporally irregular, with the background field (i.e., an earlier forecast) to create an internal grid-based representation of the atmospheric state. The significant difference in forecast skill between CloudCast and MEPS may suggest a notable discrepancy between cloudiness in the model space and the observation space. CloudCast achieves a MAESS value of 0.83 at 0-hour, a limitation due to its lower horizontal resolution. Although forecast quality decreases with longer lead times, CloudCast’s decline is less steep than that of the persistence model. Overall, CloudCast outperforms all other models for all lead times except at 0-hour. Panel b) presents the MAESS scores for CloudCast across each distinct season. Spring consistently has the highest scores over all lead times, with a difference of about 0.1 compared to other seasons. Spring is characterised by dynamic small-scale processes initiated by increased solar radiation, leading to more variable cloud cover. Winter and summer exhibit very similar performance, which is somewhat surprising given their contrasting characteristics: winter is dominated by frequent low-pressure systems and limited short wave radiation, while summer, with its abundant daylight, promotes convective activity, leading to daytime cloud development and clearer conditions at night. The lowest scores are observed in autumn, which, like winter, tends to have more overcast conditions. Panel c) displays the MAESS for intra-hour predictions over all seasons, using EXIM as the benchmark model. At 0-hour CloudCast displays the same bias observed in panel a), attributed to its coarser horizontal resolution. EXIM’s performance declines noticeably by the 15-minute mark, where it scores lower than the persistence model. This downward trend continues, albeit at a slower rate, throughout the forecast period. CloudCast consistently outperforms both EXIM and the persistence model across all forecast lead times, with a margin of approximately 0.1. At all lead times, EXIM underperforms compared to the persistence forecast — likely because it tends to shift clouds in the wrong direction. Given that cloud movement is generally slow, even a static persistence forecast remains a strong baseline.

\begin{figure}
\center
\includegraphics[width=19pc]{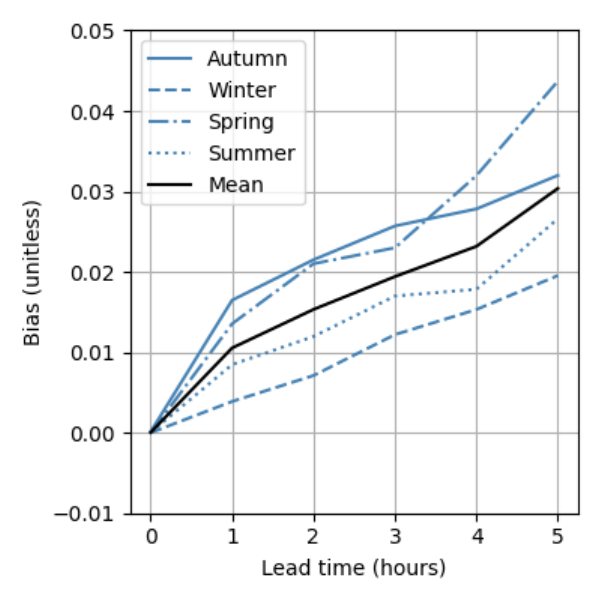}
\caption{Mean bias for each lead time}\label{f-bias-time}
\end{figure}

Figure \ref{f-bias-time} shows the mean bias of CloudCast, defined as the difference between forecast and observation, as a function of lead time. CloudCast tends to overestimate cloud cover, reflected by a positive bias. The bias increases linearly across all seasons, becoming more pronounced with longer lead times, reaching a mean bias of 0.03 by forecast hour 5. Winter and summer generally exhibit less bias compared to spring and autumn. The positive bias may stem from the model’s difficulty in capturing cloud dissipation processes over time. Additionally, as forecast uncertainty increases, the model expresses this as a blurring effect, leading to higher bias. This occurs because cloud cover data is typically concentrated near zero and one, meaning that a gradual shift toward intermediate values inflates the mean bias.

\begin{figure}
\center
\includegraphics[width=39pc]{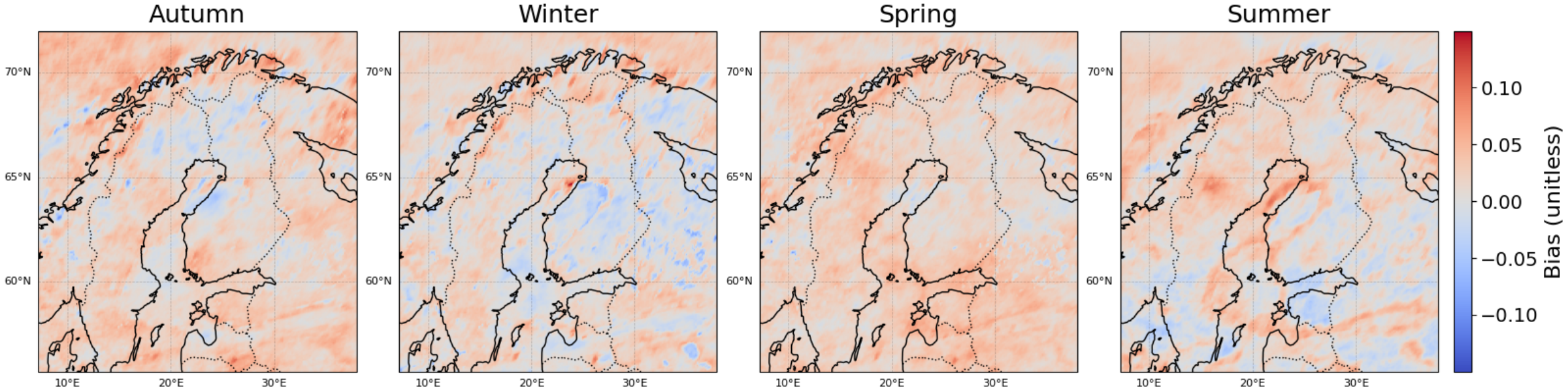}
\caption{Spatial distribution of mean bias for each season}\label{f-bias-map}
\end{figure}

In figure \ref{f-bias-map} the seasonal spatial distribution of the bias is displayed on a map centered around Finland and its neighbouring areas. Positive bias is the dominant pattern across all four seasons, especially in autumn and winter. Negative bias becomes more prevalent in spring and summer, with the summer showing the most pronounced balance between positive and negative biases. The northern and western regions consistently show more overestimation, while the eastern and southern parts exhibit more underestimation, particularly in spring and summer. Topography or land-sea distribution appears to have no significant influence on the bias. During winter low sun angle and snow covered ground can alter surface reflectance and emissivity, potentially affecting the data used for prediction. In summer, high radiance levels may lead to nonlinear cloud formation and dissipation processes that the model struggles to capture accurately. While the diurnal cycle is not explicitly represented in the data, they may still contribute to seasonal biases.

The Fractions Skill Score (FSS) \citep{Roberts2008} is a widely adopted metric for spatial verification of categorized events. As a neighbourhood-based method, FSS evaluates how well a forecasted event matches the observed event within a specified spatial threshold. It is commonly used to quantify the spatial scale at which the forecast remains useful or \textit{skillful} \citep{Mittermaier2010}.

\begin{table}[h]
\caption{Categorisation of the continuous cloudiness values}\label{t-categories}
\begin{center}
\begin{tabular}{ccccrrcrc}
\topline
$\#$ & $Description$ & $Cloudiness$ & $Octas$ & $Density$ \\
\midline
0 & Clear & [0, 0.0625] & 0 & 0.17 \\
1 & Partly cloudy & (0.0625, 0.5625] & 1,2,3,4 & 0.12 \\
2 & Mostly cloudy & (0.5625, 0.925] & 5,6,7 & 0.34 \\
3 & Overcast & (0.925, 1] & 8 & 0.49 \\ 
\botline
\end{tabular}
\end{center}
\end{table}

CloudCast forecasts are divided into four cloudiness categories, as outlined in Table \ref{t-categories}. The categorization is done by simply dividing the values by 12.5 and rounding, converting the data into octas. These fractional categories are then grouped into two broader classifications: partly cloudy (octas 1, 2, 3, 4) and mostly cloudy (octas 5, 6, 7). The density columns indicate how much of the total data volume is distributed across each category.

\begin{figure}
\center
\includegraphics[width=39pc]{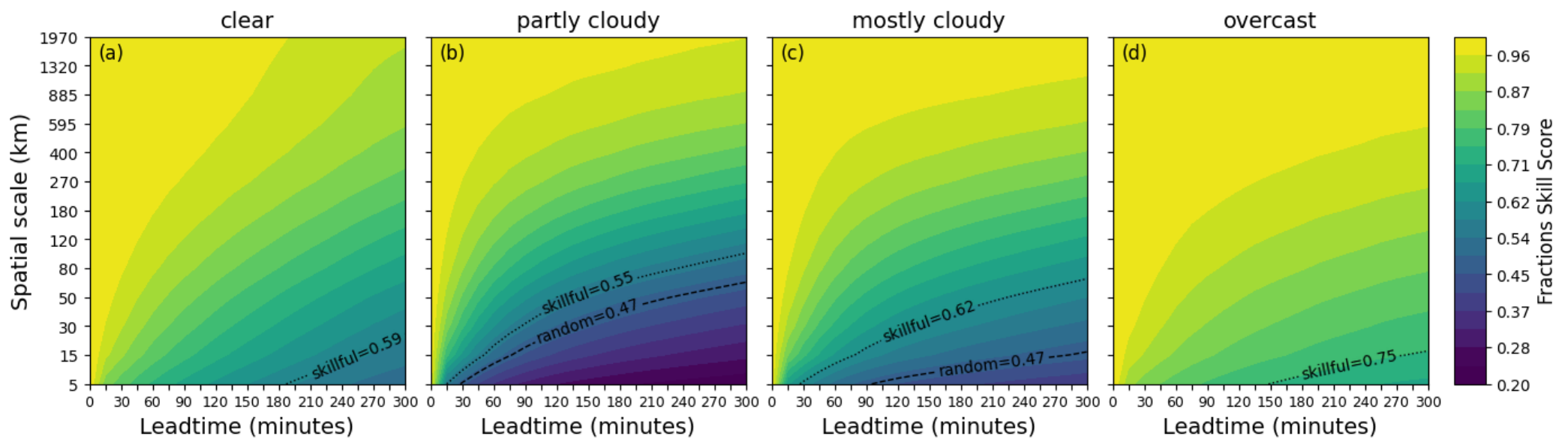}
\caption{Fractions Skill Score over all four cloudiness categories. Limits for skillful and random forecasts are drawn. A skillful forecast is better than a reference forecast; in the case of FSS a reference forecast is a forecast that assigns the overall observed frequency of the event uniformly across the entire domain. In the case of the random forecast the frequencies are assigned based on the climatology.
}\label{f-fss}
\end{figure}

The four panels in Figure \ref{f-fss} display the Fractions Skill Score (FSS) for the clear, partly cloudy, mostly cloudy, and overcast categories. Scores are calculated for all seasons, with forecasts made in 15-minute increments. The threshold for a skillful forecast, referred to as “uniform” by \citet{Roberts2008}, is represented by a dashed line. Panels b) and c) also include a second line indicating the threshold for a random forecast. In the clear sky category, CloudCast remains skillful for up to three hours, even at the smallest spatial scales. As the scale increases to 30 km, the forecast remains skillful for up to five hours. At the widest scale of 1970 km, the FSS reaches a perfect score of 1 for up to three hours. Due to computational constraints, the FSS was not calculated for the entire domain (2300 km), but based on current data, it is likely the FSS would fall short of 1 at five hours, suggesting increased bias at longer lead times. Panels b) and c) show the FSS for the partly cloudy and mostly cloudy categories, where the forecasts are less accurate compared to clear skies. For partly cloudy conditions, the forecast is skillful up to three hours at a spatial scale of about 70 km, and for mostly cloudy, up to 40 km. However, after 30 minutes for partly cloudy and 90 minutes for mostly cloudy, predictions fall below the random forecast threshold at the smallest scales. In the overcast category, the skillful range starts at around 150 minutes for the 5 km scale, reaching three hours at the 10 km scale. The FSS reaches a perfect value for the longest forecast at scales around 400 km, indicating less bias for this category compared to clear skies. Overall the FSS results indicate that CloudCast provides the most skillful forecasts for clear skies and overcast conditions, while performance degrades more quickly for partly and mostly cloudy cases. The model retains skill for up to three hours in fine-scale forecasts and up to five hours at larger scales, with the strongest performance in stable cloud regimes. However, skill drops rapidly for mixed cloud conditions, aligning with the increasing bias seen in Figure 6. This suggests that uncertainty in cloud evolution limits forecast accuracy. 

\begin{figure}
\begin{center}
\includegraphics[width=39pc]{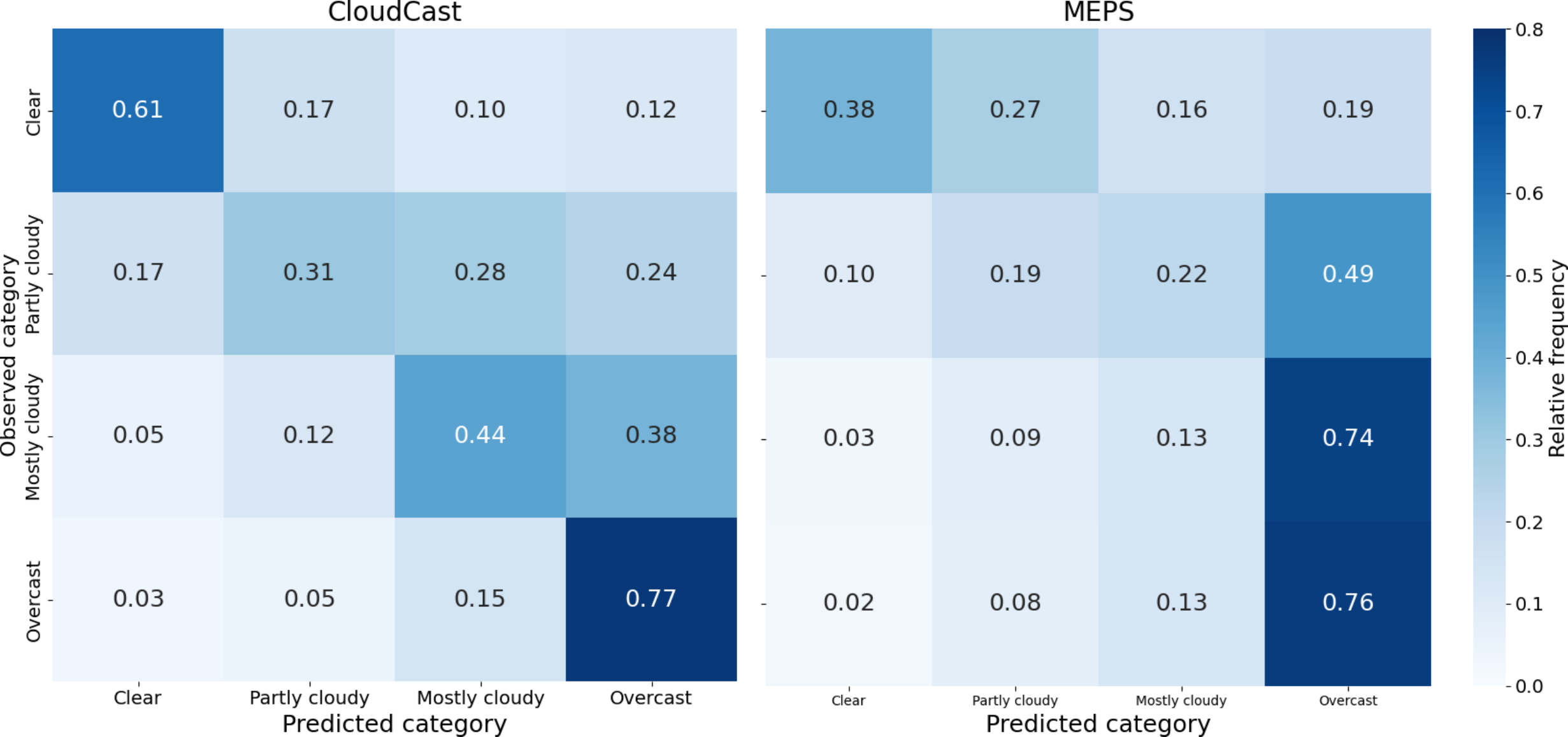}
\caption{Confusion matrix for the different cloud categories, left CloudCast and right MEPS. Entries on the diagonal indicate correct predictions.}\label{f-cm}
\end{center}
\end{figure}

Figure \ref{f-cm} displays the row-normalized confusion matrices for different cloudiness categories, comparing CloudCast (left) and MEPS (right). For CloudCast, the diagonal is well-represented, indicating a strong alignment between forecasted and observed categories. Specifically, 77\% of clear sky conditions and 61\% of overcast conditions were accurately predicted. However, there is a slight tendency to overpredict mostly cloudy as overcast (38\%). The largest discrepancy appears in the partly cloudy category, where only 31\% of forecasts were accurate, partly due to the limited data available for this category. Multi-category errors, represented in the corners of the matrix, are minimal. MEPS, on the other hand, struggles in clear, partly cloudy, and mostly cloudy conditions. Overall, CloudCast provides a more balanced prediction across categories, while MEPS struggles particularly with clear and partly cloudy conditions.

\subsection{Case studies}

\begin{figure}
\begin{center}
\includegraphics[width=39pc]{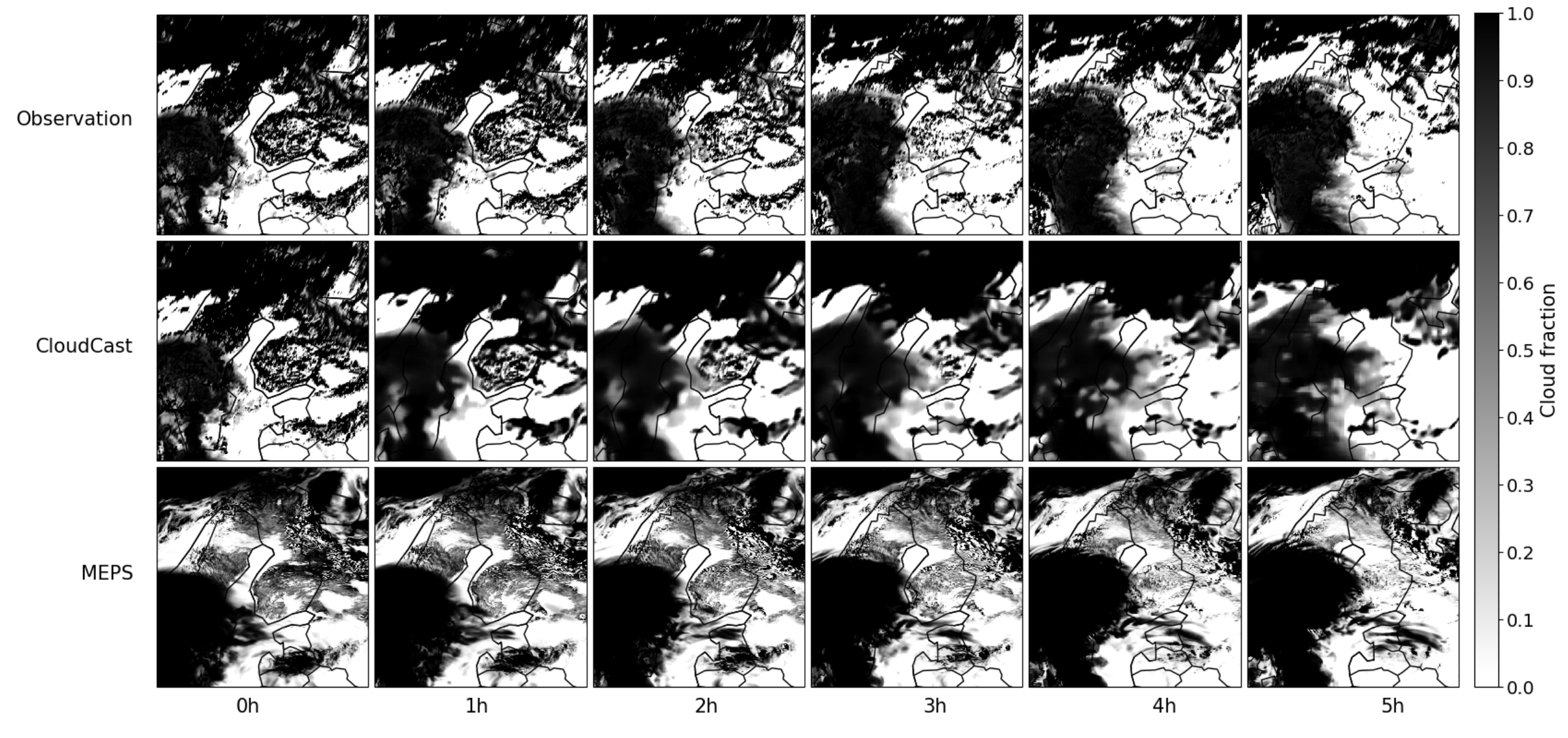}
\caption{Five hour cloud cover forecast initialized at July 10 2024 1200 UTC. A typical summertime scenario is illustrated, where  Finland is covered by convection-initiated, small-scale cloud patterns (large cumulus or stratocumulus). As the day progresses, the small-scale clouds gradually dissipate, while a frontal-based system advances towards northeast, covering the western part of Finland by the end of the forecast period. Top row is the observed value, middle row CloudCast forecast and bottom row MEPS forecast.}\label{f-cs-1}
\end{center}
\end{figure}

Three case studies are presented, with the first shown in Figure \ref{f-cs-1}. The figure includes the ground truth in the top row, the CloudCast forecast in the middle row, and the MEPS forecast in the bottom row. A typical summertime scenario is illustrated, where mainland Finland is covered by convection-initiated, small-scale cloud patterns (large cumulus or stratocumulus). A larger, frontal-based cloud system approaches from the west. As the day progresses, the cumulus clouds, formed by shortwave radiation, gradually dissipate, while the frontal system advances northeast, covering the western part of Finland by the end of the forecast period. CloudCast successfully captures the dissipation of clouds over Finland and accurately predicts the movement of the frontal system. However, the west-east oriented cloud area over the Baltics decays too slowly. As time advances, the forecast becomes smoother, and scattered clouds in the north coalesce. MEPS, in contrast, presents more fine-grained cloud structures and correctly predicts the decay trend of the clouds, although the dissipation is not strong enough. While MEPS accurately aligns the frontal system’s movement with the ground truth, it fails to decay the cloud area over the Baltics.

\begin{figure}
\begin{center}
\includegraphics[width=39pc]{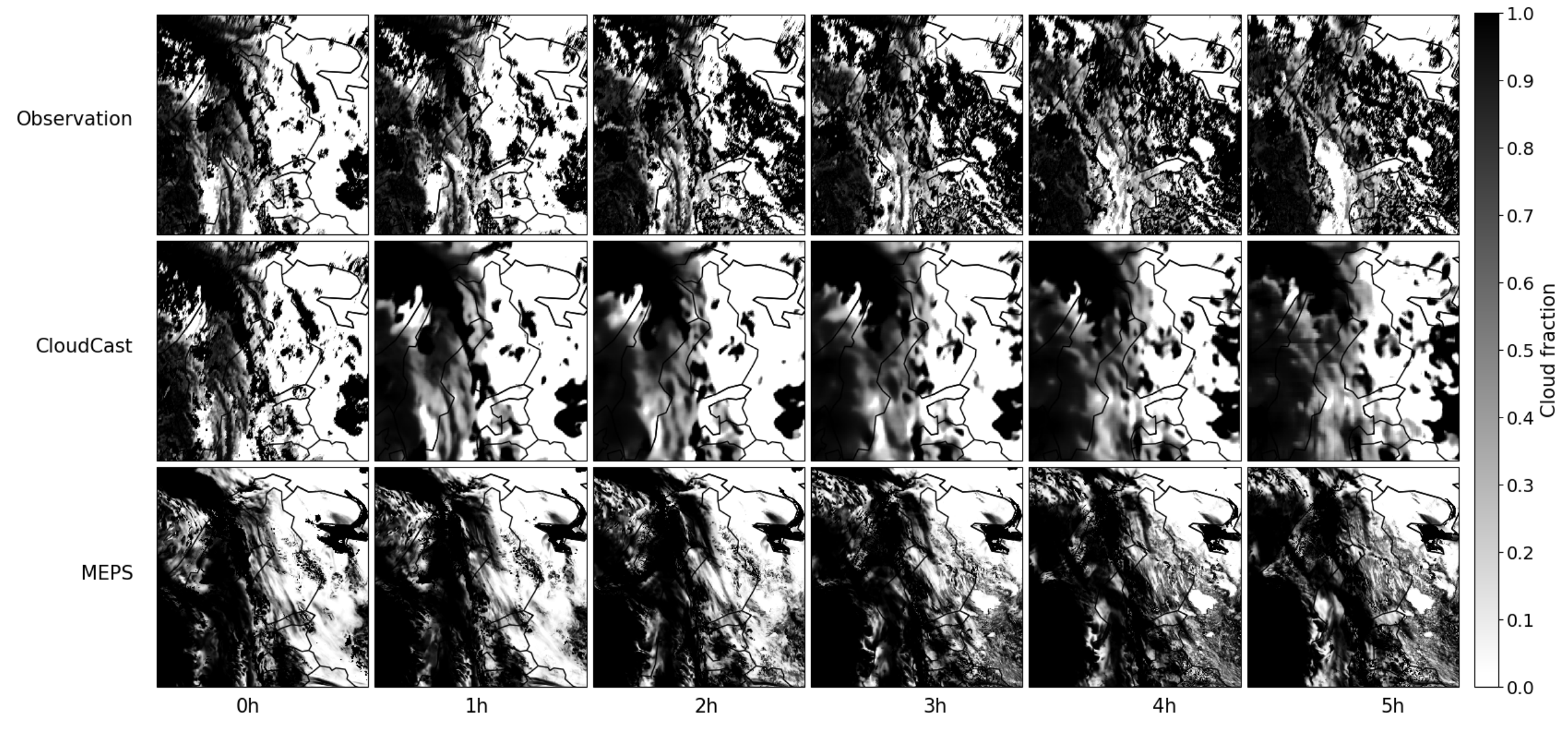}
\caption{Five hour cloud cover forecast initialized at August 9 2024 0600 UTC. The western part of the domain is initially covered by clouds, while the eastern region experiences clearer skies. Gradually clouds begin to cluster over eastern Finland, eventually spreading to cover the entire country. Top row is the observed value, middle row CloudCast forecast and bottom row MEPS forecast.}\label{f-cs-2}
\end{center}
\end{figure}

Figure \ref{f-cs-2} depicts a scenario where the western part of the domain is initially covered by clouds, while the eastern region experiences clearer skies. As the day progresses, clouds begin to cluster over eastern Finland, eventually spreading to cover the entire country. In the early hours, CloudCast captures the overall structure of the cloud cover fairly well; however, as time progresses, the forecast becomes smoother, with dense cloud regions appearing more dispersed. While CloudCast accurately generates new clouds to the correct locations later in the forecast, there is a noticeable temporal lag, and the extent of the new cloud formation is underestimated. In the western part of the domain, CloudCast successfully predicts the gradual closure of a gap in the cloud cover. MEPS forecast shows consistent cloud coverage and accurately predicts the cloud formation over eastern Finland. However, in the initial stages, it incorrectly places clouds over the White Sea, where they persist throughout the forecast period.

\begin{figure}
\begin{center}
\includegraphics[width=39pc]{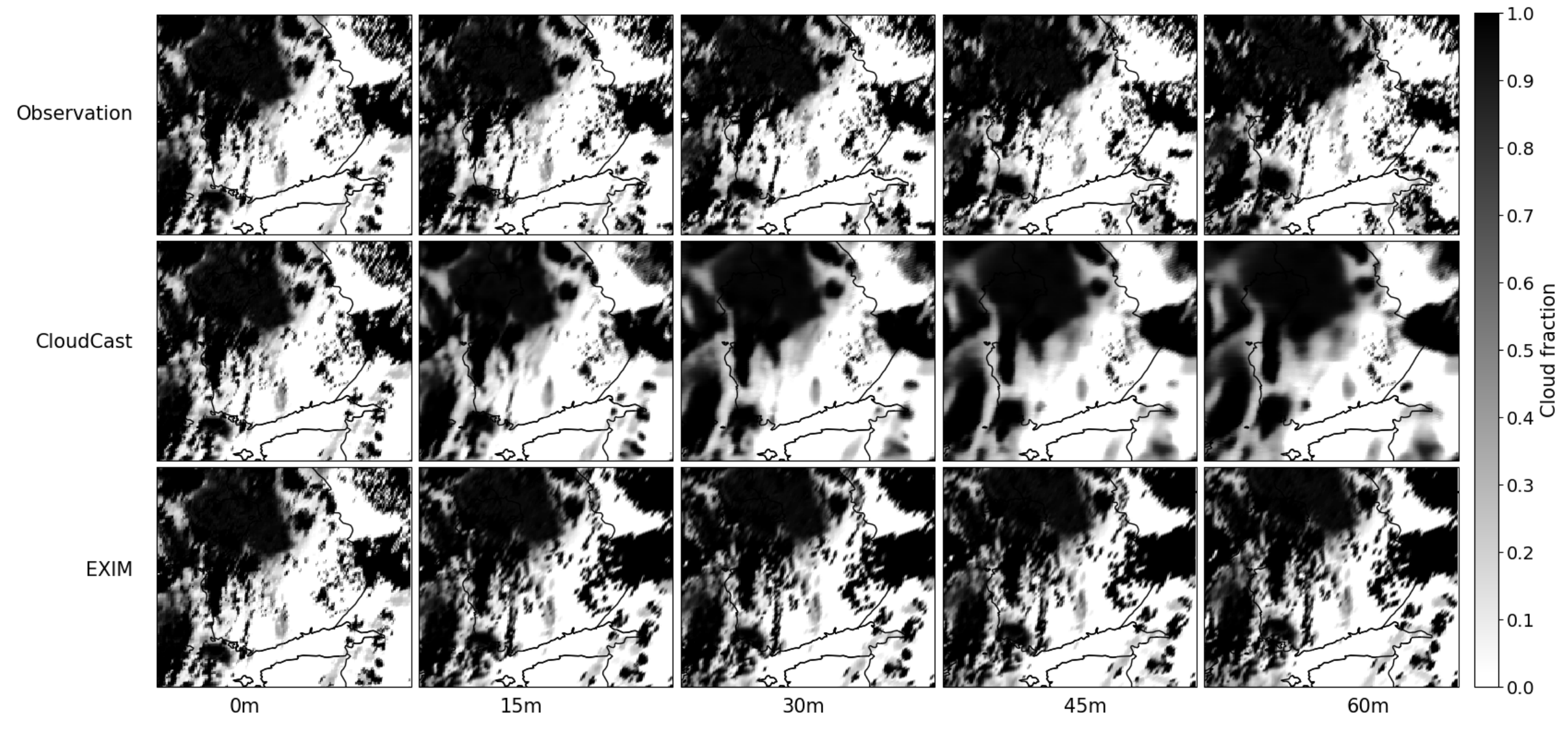}
\caption{Intra-hour cloud cover forecast initialized on August 30 2024 1100 UTC. A frontal cloud covers the western half of the domain, while Southern Finland remains under clear skies. Cloud development is minimal, with a single larger cloud element over Eastern Finland gradually dissipating. Top row is the observed value, middle row CloudCast forecast and bottom row EXIM forecast.}\label{f-cs-3}
\end{center}
\end{figure}

The third case study is presented in Figure \ref{f-cs-3}. The intra-hour forecasts from CloudCast and EXIM are shown in the middle and bottom rows, respectively. Over the span of 60 minutes, little change is observed, largely due to the lack of small-scale cloud evolution in the ground truth. A frontal cloud covers the western half of the domain, while Southern Finland remains under clear skies. Cloud development is minimal, with a single larger cloud element over Eastern Finland gradually dissipating. CloudCast captures the large-scale structure of the cloud cover but produces forecasts that appear blurry. As in the first case study, CloudCast tends to underpredict areas of overcast cloud. EXIM, on the other hand, retains small-scale features but lags in cloud development and fails to capture the full extent of cloud cover, particularly in the longer lead times.

\subsection{Operational application}

CloudCast has been used to produce the total cloud cover forecast to FMI's operational nowcasting system \citep{Hieta2021} since January 2023. A five hour forecast is produced every fifteen minutes, and the last two hours of the forecast are used to blend it with NWP model forecast. The input effective cloudiness data is preprocessed to reduce the defects inherent in the data itself. However, since the operational forecast began after the model training was completed, we have not directly quantified the improvement in forecast skill relative to the verification results in Figure 5. Additionally, verification is challenging because the ground truth data is also preprocessed, meaning that the improved quality is not necessarily reflected in simple objective scores like MAE but requires either verification against third-party observations or subjective evaluation by domain experts.

The operational forecasting process, including reading input data from object storage, evaluating the model, downscaling the results to 2.5km resolution, converting the output data into GRIB file format, and writing the output data takes around 10 seconds using an Nvidia MIG 24G GPU.

\section{Summary and discussion}

This study introduced CloudCast, a convolutional neural network based on the U-Net architecture, developed to predict total cloud cover. Using numerical experiments with relevant benchmark datasets, we find that CloudCast provides more accurate forecasts than competing NWP models and observation extrapolators. CloudCast outperforms the benchmark models in the nowcasting range of 0-5 hours. The integration of CloudCast into the Finnish Meteorological Institute’s (FMI) operational weather forecasting system has enhanced the quality of forecasts, benefiting FMI’s private and public sector clients.

CloudCast is capable of generating forecasts up to five hours ahead, though the later forecast periods are affected by increasing blurring. The observed blurring in the forecasts is consistent with prior research, which has highlighted challenges in forecasting small-scale cloud features using convolutional networks. This blurring is likely a limitation of the model, driven by the inherent unpredictability of atmospheric conditions and cloud dynamics. Tweaking hyperparameters or adjusting the depth of the network may offer minor improvements, but more recent studies have shown that more complex architectures, such as transformers, perform better when sufficient training data are available. Increasing the horizontal resolution of CloudCast could potentially sharpen the forecasts, but at the cost of introducing errors growing over time, as small-scale atmospheric dynamics are harder to predict. Another key limitation is the resolution of the ground truth data, which, at 4 km, is insufficient to fully capture summer-time convection, such as cumulus clouds. Improving the accuracy and resolution of cloud cover observations would likely enhance the training and performance of data-driven models like CloudCast.

A typical observation extrapolator cannot dynamically create and dissipate clouds. While CloudCast demonstrates some ability to generate new clouds, as seen in the case studies, the formations are often too weak and exhibit a temporal lag. Extending the skillful forecast period beyond three hours would require a stronger tendency for cloud creation and dissipation, which would likely necessitate a deeper understanding of cloud dynamics and the inclusion of additional weather-related parameters in the initial conditions, such as wind or moisture content.

CloudCast has only been tested over Northern Europe, but in principle, the same neural network architecture could applied to other regions with similar meteorological conditions, provided that sufficient cloud cover training data is available. Its transferability across different geostationary satellites is not a concern, as CloudCast does not use direct multispectral observations but rather post-processed cloud cover data. However, key limitations exist in generalizing these results. First, the ground truth data in the current domain contains artifacts due to its location at the edge of the satellite’s detectable area. These distortions, present during training, may not apply in regions closer to the satellite nadir, potentially affecting model performance. Second, atmospheric conditions play a crucial role: in Northern Europe, the dominant flow is from the southwest, and the model’s ability to generalize may be reduced in regions with significantly different prevailing wind patterns and cloud dynamics.

Future research should focus on extending CloudCast’s forecast horizon and evaluating its long-term performance. Achieving this will likely require more complex network architectures with greater capacity to learn relationships within the data. Additionally, incorporating higher-resolution cloud cover analysis data, such as from satellite observations, could help better represent small-scale cloud dynamics, further improving forecast accuracy.

\clearpage

\acknowledgments

We thank Marko Laine from the Finnish Meteorological Institute and Alex Jung from Aalto University for proofreading this manuscript and providing valuable input, which helped improve the final version.

\datastatement

Code, model weights and example data available at \url{https://github.com/fmidev/cloudcast}.

\bibliographystyle{ametsocV6}
\bibliography{references}

\end{document}